\documentclass[final,5p,times,twocolumn]{elsarticle}

\usepackage{lineno,hyperref,xcolor}
\usepackage{amsmath} 
\modulolinenumbers[5]
\biboptions{square,comma,sort&compress}
\journal{}

\begin{document}

\begin{frontmatter}

\title{Multiscale Modeling of Abnormal Grain Growth: Role of Solute Segregation and Grain Boundary Character}
%\title{Multiscale Modeling of Abnormal Grain Growth in $\alpha$-Fe: Role of Solute Segregation and Grain Boundary Character}

% \tnotetext[mytitlenote]{Fully documented templates are available in the elsarticle package on \href{http://www.ctan.org/tex-archive/macros/latex/contrib/elsarticle}{CTAN}.}

% \address{Radarweg 29, Amsterdam}
% \fntext[myfootnote]{Since 1880.}

\author[a]{Albert Linda}
\author[a]{Rajdip Mukherjee*\corref{cor}}
\ead{rajdipm@iitk.ac.in}
\author[a]{Somnath Bhowmick*\corref{cor}}
\ead{bsomnath@iitk.ac.in}
\cortext[cor1]{Corresponding Authors}
\address[a]{Department of Materials Science and Engineering, Indian Institute of
Technology, Kanpur, Kanpur-208016, UP, India}

\begin{abstract}
Abnormal grain growth (AGG) influences the properties of polycrystalline materials; however, the underlying mechanisms, particularly the role of solute segregation at the grain boundary (GB), are difficult to quantify precisely. This study demonstrates a multiscale framework that integrates atomic-scale segregation energetics (using density functional theory) with mesoscale grain growth dynamics (using phase-field model) to investigate AGG, using $\alpha$-Fe as an example system. Multisite segregation energies are calculated for symmetric tilt grain boundaries (STGBs) along the $\langle 110 \rangle$ axis for nine different solutes (Co, Cr, Mn, Mo, Nb, Ni, Ti, W, and V), encompassing three different types of coincident site lattice (CSL) boundaries: $\sum 3 (11\bar{2})$, $\sum 9 (\bar{2}21)$, and $\sum 3 (\bar{1}11)$. The model takes into account the effect of solute drag on GB mobility, estimated using a bulk solute concentration of 0.1 at\%. The results demonstrate that AGG originates due to GB anisotropy, the extent of which largely depends on the type of solute atom present. Such a complex dependence necessitates using a multiscale model to understand AGG comprehensively.  In general, low-energy $\Sigma 3$ boundaries are found to have higher mobility and show preferential growth for most of the solutes, other than Co. The study reveals how the distribution of GB types significantly influences AGG. When 10-30\% of the GBs are high-mobility type, crown-like morphologies are observed, leading to AGG. These findings underscore the critical role of GB chemistry and crystallography in governing AGG, and the model can be generalized to provide a predictive framework for controlling grain growth through strategic solute design in advanced alloys.
%Abnormal grain growth (AGG) influences the properties of polycrystalline materials; however, the underlying mechanisms, particularly the roles of solute segregation and grain boundary (GB) character, are not yet completely understood. This study employs a multiscale framework integrating atomic-scale segregation energetics  density functional theory) With mesoscale grain growth dynamics (using phase-field model) to investigate AGG in $\alpha$-Fe. Multisite segregation energies are calculated for symmetric tilt grain boundaries (STGBs) along the $\langle 110 \rangle$ axis for nine different solutes (Co, Cr, Mn, Mo, Nb, Ni, Ti, W, and V), encompassing three different types of coincident site lattice (CSL) boundaries: $\sum 3 (11\bar{2})$,  $\sum 9 (\bar{2}21)$, and $\sum 3 (\bar{1}11)$. A bulk solute concentration 0.1 at% estimates GB mobility due to solute drag. The results demonstrate that solute segregation and GB anisotropy govern AGG behavior, with low-energy $\Sigma 3$ boundaries showing higher mobility and preferential growth. The distribution of GB types significantly influences AGG behavior. When 10-30\% of the GBs are high-mobility type, crown-like morphologies are observed, leading to AGG. These findings underscore the critical role of GB chemistry and crystallography in governing AGG and provide a predictive framework for controlling grain growth through strategic solute design in advanced alloys.
\end{abstract}


\end{frontmatter}

%\linenumbers

\section{Introduction}
The size, morphology, and distribution of grains significantly affect the characteristics of polycrystalline metals and alloys, for example, mechanical, electrical, and magnetic properties~\cite{Priester2012}. It has been observed that, under certain conditions, a small fraction of grains may grow rapidly, consuming neighboring grains in a polycrystalline material; a phenomenon known as abnormal grain growth (AGG)~\cite{Gladman1992, Najafkhani2021}. Accurately predicting the onset of AGG remains a long-standing challenge in materials science\cite{Krill2023}, as it critically affects the properties of a material. 

In ductile materials, for example, the Hall-Petch relationship suggests that larger grains exhibit reduced mechanical strength compared to smaller grains~\cite{HANSEN2004801}. Conversely, large grains may serve as critical flaws in brittle materials, reducing fracture toughness~\cite{Kingery1976}. Even in low-stress environments, a few oversized grains within a fine-grained matrix can have detrimental effects, such as the roping phenomenon observed in certain aluminum alloys being considered for automotive body panels~\cite{Ouhiba2022}. On the positive side, AGG can be beneficial in certain applications, such as promoting the growth of low-resistivity copper interconnects~\cite{Lu2014} or reducing core losses in transformers~\cite{Munetsugu1989}. Furthermore, a highly bimodal grain size distribution can improve the ductility of bulk copper samples subjected to cryogenic rolling and subsequent annealing, where recrystallization and AGG are triggered~\cite{Wang2002}. % Traditionally, it has been assumed that GB curvature is responsible for high GB velocity; however, it has been shown recently that texture plays a much more critical role \cite{Aditi2021}.

Previous studies have investigated AGG using both cellular automata~\cite{ma17010138} and phase-field modeling~\cite{KIM20083739}. In phase-field simulations, various models have been employed to explore AGG behavior. These include studies on complexion-transitioned boundaries at triple junctions, which provide driving forces for partial wetting and the formation of quadruple junctions~\cite{DE2023112451}. The role of second-phase particles in AGG has also been explored~\cite{Suwa2022, KINOSHITA2020109558}, along with the effects of non-uniform plastic deformation that creates locally high restored energy in specific normal grains~\cite{WU2020100790}, and strain-induced grain growth driven by stored deformation energy~\cite{LIU20223873}. Additionally, it has been shown that an initial distribution of large grains can effectively regulate the growth rate of abnormally large grains~\cite{ZHANG20213395, Miyoshi2022}. Anisotropic interface mobility and recrystallization-type driving forces have been identified as key contributors to AGG in some cases~\cite{KIM20083739}, while recently it has been observed that the presence of $\langle 110 \rangle$ type GB tends to show AGG~\cite{KUNDIN2020109926}. These recent studies hinted at the possibility of AGG through the combined effect of solute drag (due to segregation) and GB anisotropy.

% Despite numerous studies, there remains no conclusive evidence identifying the specific type of GBs responsible for AGG, nor has the role of solute segregation been fully understood in this context. Additionally, there is a lack of multiscale frameworks that comprehensively address these factors.

Although AGG has been widely studied, the underlying mechanisms, particularly the role of solute segregation at the grain boundary (GB), are difficult to quantify precisely. A significant challenge is the lack of multiscale modeling frameworks linking atomic-level phenomena, such as segregation energies, with grain growth behavior observed at the continuum scale. Such an integration is crucial, as solute segregation arises from atomic-scale energetics, whereas AGG is driven by mesoscale microstructural evolution.  No single computational method can adequately capture the complex interactions across these different length scales. Bridging this scale gap using a multiscale model is the key to achieving a more accurate and predictive understanding of AGG and its underlying mechanisms.

In this study, we conduct a multiscale investigation of grain growth driven by solute drag from solute segregation, combining the multisite McLean isotherm~\cite{Lejcek2010} and the Cahn~\cite{CAHN1962789}, L\"ucke, and St\"uwe~\cite{Lucke1972} (CLS) model (utilizing ab initio calculations) and an anisotropic grain growth model~\cite{PhysRevLett.101.025502} (utilizing phase-field simulations).  We consider nine different solutes (Co, Cr, Mn, Mo, Nb, Ni, Ti, W, and V) segregating along the $\langle 110 \rangle$ GB axis, as this orientation has recently been linked to AGG in iron~\cite{ALMEIDAJR202011099}. Additionally, the low-energy $\sum 3$ twin boundary for $\langle 110 \rangle$  has been reported to have the highest mobility among all high-angle coincident site lattice (CSL) boundaries~\cite{TODACARABALLO20121116, DAPHALAPURKAR201482}. Based on these insights, we select three types of high-angle symmetric tilt grain boundaries (STGBs) for our study:  $\sum 3 (\bar{1}11)$, $\sum 3 (11\bar{2})$, and $\sum 9 (\bar{2}21)$. Among these, the $\sum 3 (11\bar{2})$ boundary is low-energy, while the other two are high-energy STGBs. All three GBs were generated using the CSL approach, which accurately represents GBs at various misorientations~\cite{GARBACZ19951541}. We find low-energy $\Sigma 3$ boundaries to have higher mobility, and they show preferential growth for most of the solutes, other than Co. Our study uncovers how the distribution of GB types influences AGG; if 10-30\% of the GBs are high-mobility type, crown-like morphologies are observed, leading to AGG. 

\section{Computational Details}
%This section details the step-by-step model implementation, starting with the multisite segregation energy calculations using \textit{ab initio} methods, followed by the GB concentration determination using the McLean isotherm. Next, we employ the CLS model to derive the GB mobilities and use this data to explore anisotropic grain growth in $\alpha$-Fe due to solute segregation using phase-field simulations.
\begin{figure}
\centering 
\fcolorbox{white}{white}{\includegraphics[width=1.0\linewidth]{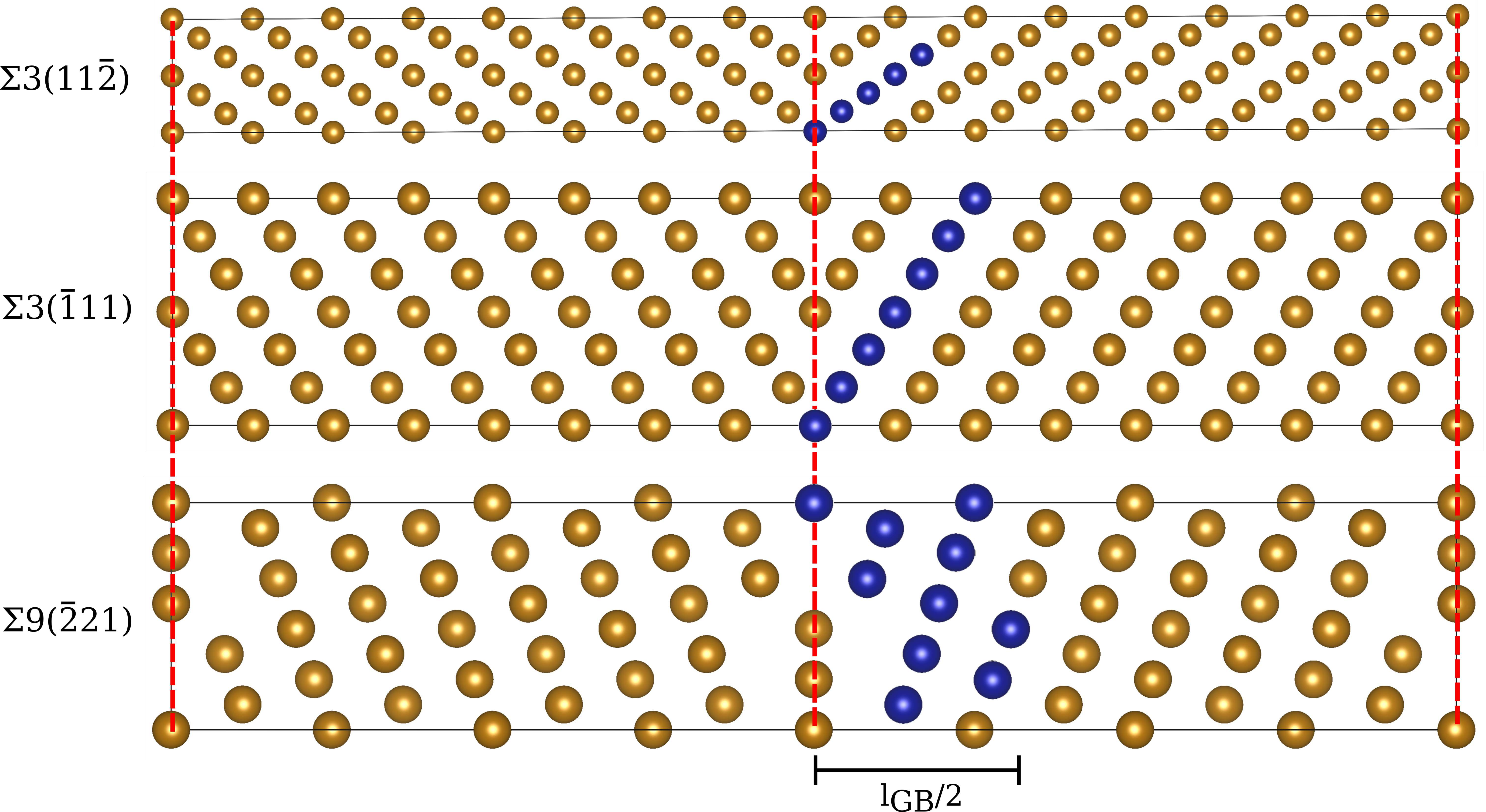}}
\caption{GB structures for STGBs, where the dotted red line denotes the center of the GB, having width $l_{GB}$. Segregation energies are calculated by placing a single solute atom at various segregation sites, highlighted in blue. The topmost structure corresponds to a twin boundary, while the remaining ones are symmetric tilt boundaries. The structures were visualized using VESTA~\cite{VESTA}.}
\label{fig:schematic}
\end{figure}

\begin{table}[]
\caption{Details of supercells used to calculate GB energy (GBE).}
\small
\begin{tabular}{|l|l|l|l|l}
\cline{1-4}
GBs                         & Dimensions & Atoms & GBE (mJ/m$^2$) &  \\ \cline{1-4}
$\sum 3 (\bar{1}11$) & \begin{tabular}[c]{@{}l@{}}$4.0 \times 4.9 \times 56.0$\\ $\alpha=\beta=\gamma=90^\circ$\end{tabular}       & 96              & 1.56         &  \\ \cline{1-4}
$\sum 3 (11\bar{2})$ & \begin{tabular}[c]{@{}l@{}}$4.0 \times 7.0 \times 39.6$\\ $\alpha=\beta=\gamma=90^\circ$\end{tabular}       & 96              & 0.43        &  \\ \cline{1-4}
$\sum 9 (\bar{2}21)$ & \begin{tabular}[c]{@{}l@{}}$4.0 \times 6.4 \times 34.2$\\ $\alpha=\beta=90,\gamma=108.4^\circ$\end{tabular} & 68              & 1.74         &  \\ \cline{1-4}
\end{tabular}

\label{table:gb_dimensions}
\end{table}
\subsection{Structure generation}
Using the coincident site lattice (CSL) approach, we create high-angle grain boundaries using \textit{aimsgb} tool~\cite{CHENG201892}. We use three distinct types of STGB: $\sum 3 (\bar{1}11)$, $\sum 3 (11\bar{2})$, and $\sum 9 (\bar{2}21)$, as shown in Figure \ref{fig:schematic}. To ensure that the atoms do not lie too close, the atoms closer than 0.3 times the first nearest neighbor distance are merged into a single atom. The details of the final structure obtained are listed in Table \ref{table:gb_dimensions}. Due to periodic boundary conditions, the resulting structures have two grain boundaries, one at the midpoint and the other at the supercell's edge. To ensure minimal interaction between these two GBs, we maintain a separation distance greater than 15 \AA~ between them.

\subsection{DFT calculation}
Atomic positions and cell vectors are relaxed to their equilibrium values, and energies are computed using the density functional theory (DFT) calculations, as implemented in the Vienna Ab initio Simulation Package (VASP)~\cite{PhysRevB.54.11169}. We take a kinetic energy cutoff of 400 eV for all calculations. We use the Methfessel-Paxton electron smearing~\cite{PhysRevB.40.3616}, with a smearing width of 0.1 eV. We fix energy convergence criteria of $10^{-6}$ eV for the electronic step and force convergence criteria of $0.01$ eV/\AA~ for the ionic step. To sample the Brillouin zone, we utilize a $12\times7\times1$,$12\times10\times1$ and $12\times8\times1$  k-point mesh for $\sum 3 (\bar{1}11)$, $\sum 3 (11\bar{2})$ and $\sum 9 (\bar{2}21)$ respectively. Since BCC iron has a magnetic ground state, we perform all calculations by considering spin polarization to account for the magnetic effect. 

\subsection{Grain boundary energy and segregation energy}
We calculate the GB energies of pure Fe (having no solute atom) using the following expression,
\begin{equation}
\sigma_{GB} = \frac{E^{GB}-E^{bulk}}{2A},
\label{eq:gb_energy}
\end{equation}
where $E^{GB}$ and $E^{bulk}$ represent the energies of the supercells (of equal size) with and without the grain boundary, respectively, and $A$ represents the grain boundary area. The factor of 2 in the denominator accounts for two GBs in the supercell [Figure~\ref{fig:schematic}]. GB energies lie within 0.43 to 1.74 mJ/m$^2$ [Table \ref{table:gb_dimensions}]. As expected, the $\sum 3 (11\bar{2})$ being the coherent GB, has the lowest energy among all three. GB energies agree well with previously reported values~\cite{MAI2022117902}, which also validates the choice of computational parameters, as described in the previous section.

Next, we use the same GB structure to assess solute segregation at the grain boundary. As shown in Figure~\ref{fig:schematic}, a single solute atom replaces an iron atom at various lattice sites. The segregation energy at the $i^{th}$ lattice site in the presence of a solute atom is given by
\begin{equation}
E^i_{seg}= \frac{(E_{solute}^{i,GB}-E_{solute}^{bulk})-(E^{GB}-E^{bulk})}{2},
\label{eqseg}
\end{equation}
where $E_{solute}^{i,GB}$ represents the energy of the supercell with a single solute substitution at the $i^{th}$ lattice site at the grain boundary. $E_{solute}^{bulk}$ represents the energy of the defect-free supercell with a single solute substitution. We take equal-sized supercells for $E_{solute}^{i,GB}$, $E_{solute}^{bulk}$, $E^{GB}$, and $E^{bulk}$ calculation.

% To explore the underlying processes driving grain boundary segregation of various elements, we analyzed how the segregation energy for each element at the grain boundary varies with the Voronoi volume of each component site, using the difference in cohesive energies of solute and bulk atom to estimate the chemical contribution to grain boundary segregation. In the phenomenological model, segregation energy generally consists of
% \begin{equation}
%     E_{seg} = E^{c}_{seg} + E^{e}_{seg}
% \end{equation}

% \begin{equation}
%     E_{seg}^rest \frac{1}{20}\left(1-\sqrt{\frac{z_{GB}}{z_{bulk}}}\right)\left[W^A N_A(10-N_A)-W^{B\rightarrow A}N_B(10-,N_B)\rig,ht]
% \end{equation}

% \begin{equation}
% \mu=\int (\varepsilon -\varepsilon_F)g_d(\varepsilon) d\varepsilon,   
% \end{equation}

% \begin{equation}
% \sigma^2=\int (\varepsilon -\varepsilon_d)^2g_d(\varepsilon)d\varepsilon.   
% \label{eqwd}
% \end{equation}

%The following expression gives the elastic segregation energy and the volume mismatch between the solute and bulk atoms. Scheiber et al. \cite{SCHEIBER201675} proposed an expression for this elastic contribution. The following expression gives the elastic segregation energy:
% \begin{equation}
% \textstyle
%     E^{e}_{seg} =-\frac{2G_A K_B(V_A -V_B)^2}{4G_A V_A + 3K_B V_A}
%     + \frac{2G_A K_B(V^{gb}_A -V_B)^2}{4G_A V_B + 3K_B V^{gb}_A} 
%     - \frac{2G_A K_A(V^{gb}_A -V_A)^2}{4G_A V_A + 3K_A V^{gb}_A} 
%     \label{eq:eseg_elastic}
% \end{equation}

\subsection{Grain boundary concentration and mobility}
\label{subsecgbconc}
The segregation energies from our \textit{ab initio} calculations can be used in Mc-Lean~\cite{Lejcek2010} isotherm to obtain the solute concentration at GB at a given temperature for a given bulk concentration. According to the McLean isotherm, the solute concentration at the GB is given by:
\begin{equation}
    C^{i}_{\text{GB}}=\frac{C_{b}\exp(-E^{i}_{\text{seg}}/k_B T)}{1-C_{b}+C_{b}\exp(-E^{i}_{\text{seg}}/k_{B} T)},
    \label{eq:gb_concentration_1}
\end{equation}
where $C^{i}_{\text{GB}}$ and $E^{i}_{\text{seg}}$ are the GB concentration and  segregation energy for site $i$.

%This model considers GB as a region of finite thickness within the crystalline surroundings, while the bulk of the crystal is infinite. Since we cannot take an infinite bulk system, we take a sufficiently large distance (greater than 15 \AA) between the GBs to avoid interaction between two adjacent boundaries. 

%Recently, it has been pointed out that the significance of the contribution of entropy in the segregation energy\cite{LEJCEK2021116597}. However, as in general the magnitude of entropy contribution is relatively small when compared to enthalpy\cite{LEJCEK2019253} and also in our case, since the bulk concentration of solute substitution considered is too low, the entropy contribution will be even more minor. This led us to only account for the enthalpy contribution for our study.

%\subsection{Grain boundary mobility}
When the GB moves under a driving force, the presence of a solute atom causes the concentration profile to move behind the GB. As a result, a net attraction exists between the solute atom and the GB, resulting in reduced GB mobility. This phenomenon is known as the solute drag effect. Reduction in GB mobility can be estimated using the continuum solute drag model, developed by Cahn, L\"ucke, and St\"uwe  (CLS model)~\cite{CAHN1962789, Lucke1972}. In this model, the solute excess ($\Gamma$) is given by the following equation\cite{SUHANE2022117473},
\begin{equation}
    \Gamma=2N_v C_{b} l_{GB} \left[\frac{k_B T}{E_0}\left(1-\exp\left(-\frac{E_0}{k_B T}\right)\right)-1\right].
    \label{eq:solute_excess_CLS}
\end{equation} 
In the above equation, $N_v$ is the number of atoms per unit volume, $l_{GB}$ is the GB width [Figure~\ref{fig:schematic}], $k_B$ is the Boltzmann constant, $T$ is the temperature, and $E_0$ is the effective segregation energy. One can also estimate solute excess ($\Gamma$) from atomistic calculations using,
\begin{equation}
    \Gamma_{\text{DFT}}=\frac{1}{A}\sum_{i=1}^{N}(C_{\text{GB}}^{i}-C_b).
    \label{eq:solute_excess_DFT}
\end{equation}
In the above expression, $C_{\text{GB}}^{i}$ is obtained from \textit{ab initio} calculations [Equation~\ref{eq:gb_concentration_1}]. Equating equation~\ref{eq:solute_excess_CLS} and \ref{eq:solute_excess_DFT}, we can obtain $E_0$ using a root finding method like Newton Raphson. Using the value of $E_0$, one can estimate the grain boundary mobility in the presence of solute segregation using the following equation~\cite{Mendelev2005, SUN2014137}, 
\begin{equation}
    M=\frac{D}{2N_v l_{GB}}\frac{1}{C_{b}}\frac{E_0}{(k_B T)^2}\left[\sinh\left(\frac{E_0}{k_B T}\right)-\frac{E_0}{k_B T}\right]^{-1}.
    \label{eq:mobility}
\end{equation}
In the above equation, the diffusivity $D$ is obtained from the literature [see Table S1, Supplementary Material (SM)]. GB width ($l_{GB}$) is found to be $\approx 1$ nm, with a small variation like 1.02, 0.94, and 0.96 nm for $\sum 3 (\bar{1}11)$, $\sum 3 (11\bar{2})$, and $\sum 9 (\bar{2}21)$, respectively. The values of $l_{GB}$ are decided based on segregation energies. Beyond the distances mentioned above, segregation energies become negligible ($\sim 2\%$ of the highest segregation energy at the GB center).

\subsection{Phase-field simulation}
We employ a phase-field model to study grain growth in Fe-X (Co, Cr, Mn, Mo, Nb, Ni, Ti, W, and V) single-phase alloys with grain boundaries having different energies and mobilities. In this study, we consider curvature-driven grain growth in the absence of elastic strain effects. The phase-field model was proposed initially by Moleans et al. ~\cite{PhysRevLett.101.025502}, which allows for anisotropy in GB energy and mobility. Such a model is suitable when mixed types of GBs are present in the material. It also ensures a constant diffuse interface width with the correct choice of parameters. The model provides significant control over the accuracy in grain growth simulations.

\begin{figure*}[h]
\centering\fcolorbox{black}{white}{\includegraphics[width=\linewidth]{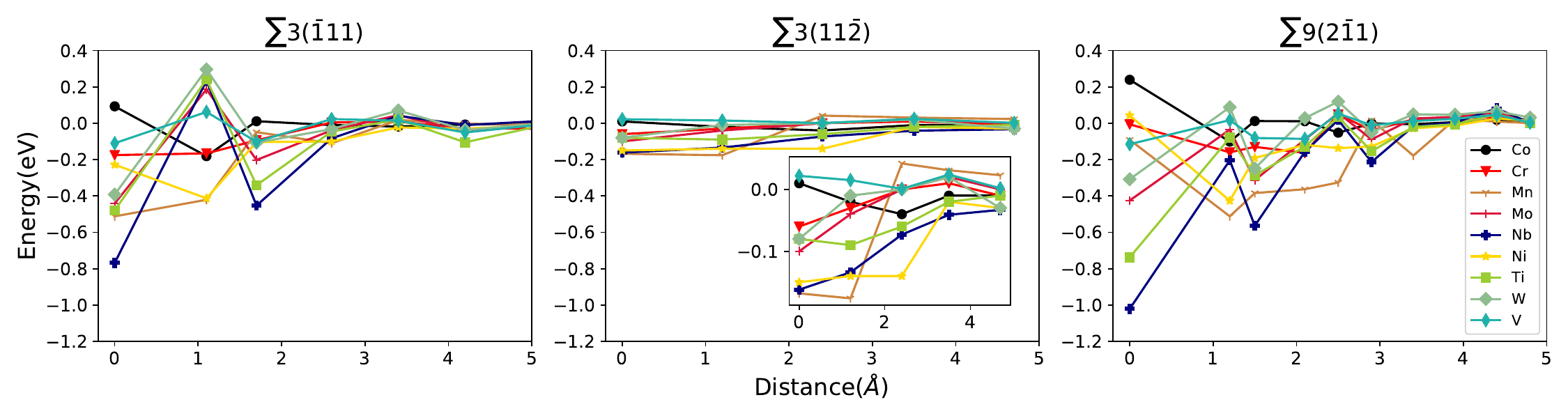}}
\caption{Site-wise segregation energies as a function of distance from the GB center. Note that, magnitude of $E_{seg}$ is relatively small in case of low energy $\Sigma 3(11\overline{2})$.}
\label{fig:eseg_sitewise}
\end{figure*}

\begin{figure*}
\fcolorbox{black}{white}{\includegraphics[width=0.55\linewidth]{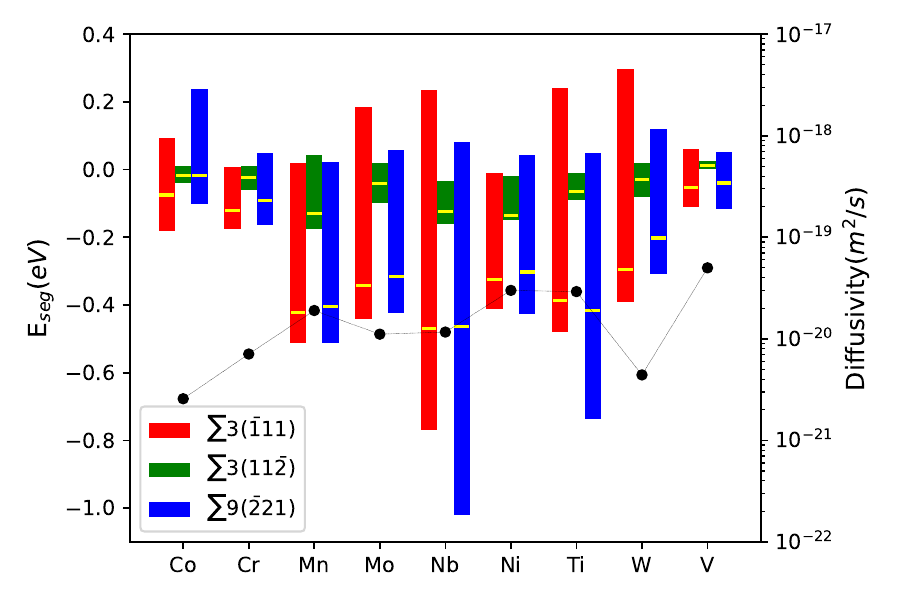}}
\fcolorbox{black}{white}{\includegraphics[width=0.46\linewidth]{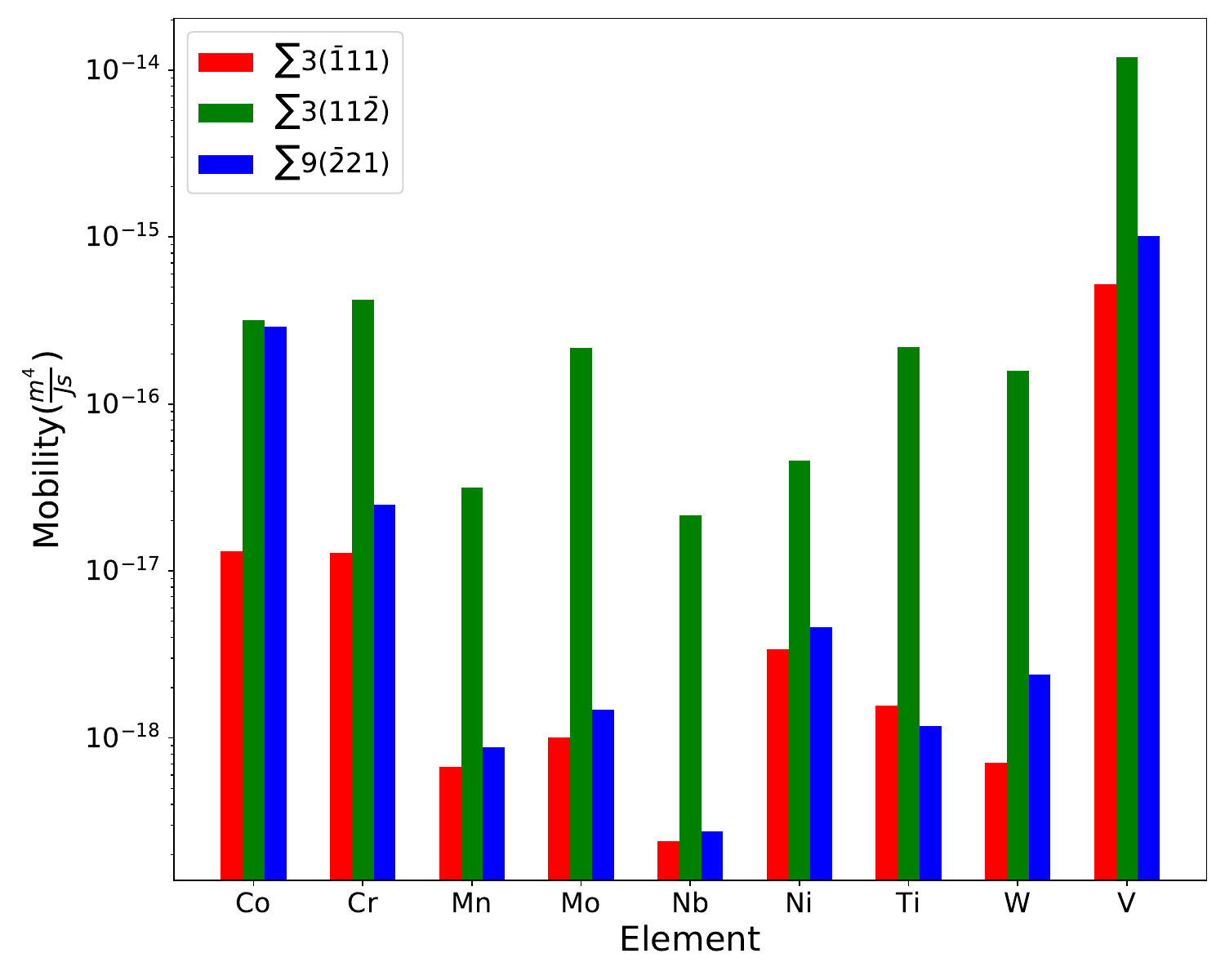}}
\caption{(Left) Bar plot illustrates the segregation energy spectrum across different segregating sites [Figure~\ref{fig:schematic}]. Yellow lines denote the value of effective segregation energy $E_0$ [Equation~\ref{eq:solute_excess_CLS}]. The circles represent the diffusivity of the respective element in $\alpha$-Fe. (Right) Grain boundary mobility [Equation~\ref{eq:mobility}] of $\alpha$-Fe with 0.1\% solute concentration at 850 K.}
\label{fig:eseg_effective_diff}
\end{figure*}

The GB mobilities estimated from the \textit{ab initio} calculations are used to study grain growth using phase-field simulations. We first define non-conserved order parameters $\eta_1$, $\eta_2$...$\eta_Q$, which account for the different grain orientations. The evolution of the order parameter with time is governed by the Allan-Chan equation,
\begin{equation}
    \frac{\partial \eta_{i}}{\partial t} = -L(\theta)\frac{\delta F}{\delta \eta_{i}}, i = 1,2...Q 
\end{equation}
where $L$ is the kinetic coefficient. The free energy functional ($F$) has the following form, 
\begin{equation}
    F = \int_V\left[f( \eta_{1}, \eta_{2},...\eta_{Q}) + \frac{\kappa(\theta)}{2}\sum_{i=1}^{Q}(\Delta \eta_{i})^2\right]dV.
\end{equation}
In the above equation, $\kappa$ is the gradient energy coefficient, and $f$ is the local free energy density, which has the following form,
\begin{equation}
    f(\eta_{1}, \eta_{2},...\eta_{Q}) = \sum_{i=1}^{Q}\left(-\frac{1}{2}\alpha \eta_{i}^2 + \frac{1}{4} \beta \eta_{i}^4\right) + \gamma(\theta) \sum_{i=1}^{Q} \sum_{j>i}\eta_{i}^{2}\eta_{j}^{2} + \frac{1}{4}.
\end{equation}
Since we consider only high-angle GBs in this work, the curvature effect dominates the grain growth kinetics~\cite{KAZARYAN20022491, HIROUCHI2012474, KIRCH20084998}, and the inclination effect can be neglected~\cite{MOELANS2022110592}. The following expression gives the parameters at each grid point required in the model. 
\begin{equation}
\scriptsize
    L(\theta) = \frac{\sum_{i=1}^{Q}\sum_{j>i}^{Q}L_{i,j}\eta^2_{i}\eta^2_{j}}{\sum_{i=1}^{Q}\sum_{j>i}^{Q}\eta^2_{i}\eta^2_{j}}; \gamma(\theta) = \frac{\sum_{i=1}^{Q}\sum_{j>i}^{Q}\gamma_{i,j}\eta^2_{i}\eta^2_{j}}{\sum_{i=1}^{Q}\sum_{j>i}^{Q}\eta^2_{i}\eta^2_{j}}; \kappa(\theta) = \frac{\sum_{i=1}^{Q}\sum_{j>i}^{Q}\kappa_{i,j}\eta^2_{i}\eta^2_{j}}{\sum_{i=1}^{Q}\sum_{j>i}^{Q}\eta^2_{i}\eta^2_{j}}.
\end{equation}
In the above expressions, $L_{ij}$, $\gamma_{ij}$, and  $\kappa_{ij}$ are values for the grain boundary between a grain with orientation $i$ and one with orientation $j$. We estimate the values of $L(\theta)$, $\kappa(\theta)$, and $\gamma(\theta)$ from the above equations using an iterative procedure~\cite{PhysRevB.78.024113}. Finally, we solve the phase field equation by discretizing it using the finite difference scheme.

Note that the presence of many grains requires a large number of order parameters, making the study of microstructure evolution computationally very expensive. To reduce the simulation time, we use the Active Parameter Tracking (APT) algorithm~\cite{Vedantam2006}, which has been successfully used to accelerate the study of thermal grooving during grain growth~\cite{VERMA2023119393}. In APT, one first prepares a list of active parameters having $\eta$ greater than a threshold value. For our case, we take the threshold to be $10^{-6}$. We only perform calculations over those grid points in the active parameter list. This method allows for solving fewer points, mainly in the GB vicinity, making the model fast. Also, as the area of GBs decreases during a later time step, it progresses even faster due to a reduction in GB area. %To further accelerate the calculation to perform computation for a large number of grains, we parallelized our calculation using the OpenMP library. 

\section{Results and Discussion}
\subsection{Effective Segregation Energy}
\label{subsecSE}
We evaluate segregation energies in the presence of nine different transition elements, namely Co, Cr, Mn, Mo, Nb, Ni, Ti, W, and V, for all three STGBs and at various substitutional sites [Figure~\ref{fig:schematic}]. A negative segregation energy indicates the tendency for the solute atom to segregate at the boundaries, with a higher magnitude indicating higher segregation tendency~\cite{RAZUMOVSKIY2015369}. The segregation energies near the GB centers are generally large in magnitude and negative in sign; these values decrease as one moves away from the GB center [Figure~\ref{fig:eseg_sitewise}]. Such a trend confirms that the GBs are diffused, and their influence decreases gradually away from the center~\cite{LEJCEK201783}. 

We calculate the effective segregation energy ($E_0$), starting with the site-wise segregation energies $E^i_{seg}$ [Equation~\ref{eqseg}]. We first calculate site-wise solute concentration $C^i_{GB}$ [Equation~\ref{eq:gb_concentration_1}], which is used to estimate the solute excess $\Gamma_{DFT}$ [Equation~\ref{eq:solute_excess_DFT}]. Equating this quantity with the solute excess $\Gamma$ from the CLS model [Equation~\ref{eq:solute_excess_CLS}], we obtain the effective segregation energy $E_0$. These values, along with the segregation energy spectrum, are illustrated in the left panel of Figure~\ref{fig:eseg_effective_diff}. The segregation energy spectrum is narrow, and $E_0$ values are lower in magnitude for $\sum 3(11\bar{2})$ GB. On the other hand, the segregation energy spectrum is wide, and $E_0$ values are higher in magnitude for $\sum 3 (\bar{1}11)$ and $\sum 3 (11\bar{2})$ GBs.
%We also show the segregation energy spectrum for each solute in Figure~\ref {fig:eseg_effective_diff}. Although most segregation energies are negative, we also observe a few small positive values at specific locations. Interestingly, the segregation energies depend on the type of GB. For the low energy boundary $\sum 3(11\bar{2})$, the segregation energies are smaller in magnitude compared to the other two types of GBs. 

This trend can be understood in terms of Voronoi volume. At lattice sites near the GB center, the Voronoi volume is closer to its bulk value in the case of $\sum 3(11\bar{2})$. On the other hand, the Voronoi volume at lattice sites near the GB center in the case of $\sum 3 (\bar{1}11)$ and $\sum 9 (\bar{2}21)$ differs significantly from the bulk Voronoi volume. A more considerable Voronoi volume difference between the bulk and GB region increases the elastic contribution to the segregation energy, which explains its higher magnitude for $\sum 3 (\bar{1}11)$ and $\sum 9 (\bar{2}21)$, compared to $\sum 3(11\bar{2})$ [see Figure S1, SM]. 

As the magnitude of segregation energy is less in low energy $\sum 3(11\bar{2})$ boundary [Figure~\ref{fig:eseg_effective_diff}, left panel], solute atoms have less tendency to segregate at this boundary. As a result, solute segregation at the GB in the case of low energy $\sum 3(11\bar{2})$ boundary is less, irrespective of the type of solute atom. The difference in solute segregation at low vs. high energy GB is more prominent at elevated temperatures [see Figure S2, SM]. Among the solute atoms studied in this work, Cr, Co, and V have a lower segregation tendency than the other elements.

\subsection{GB mobility}
\begin{figure}
\centering 
\fcolorbox{black}{white}{\includegraphics[width=1.0\linewidth]{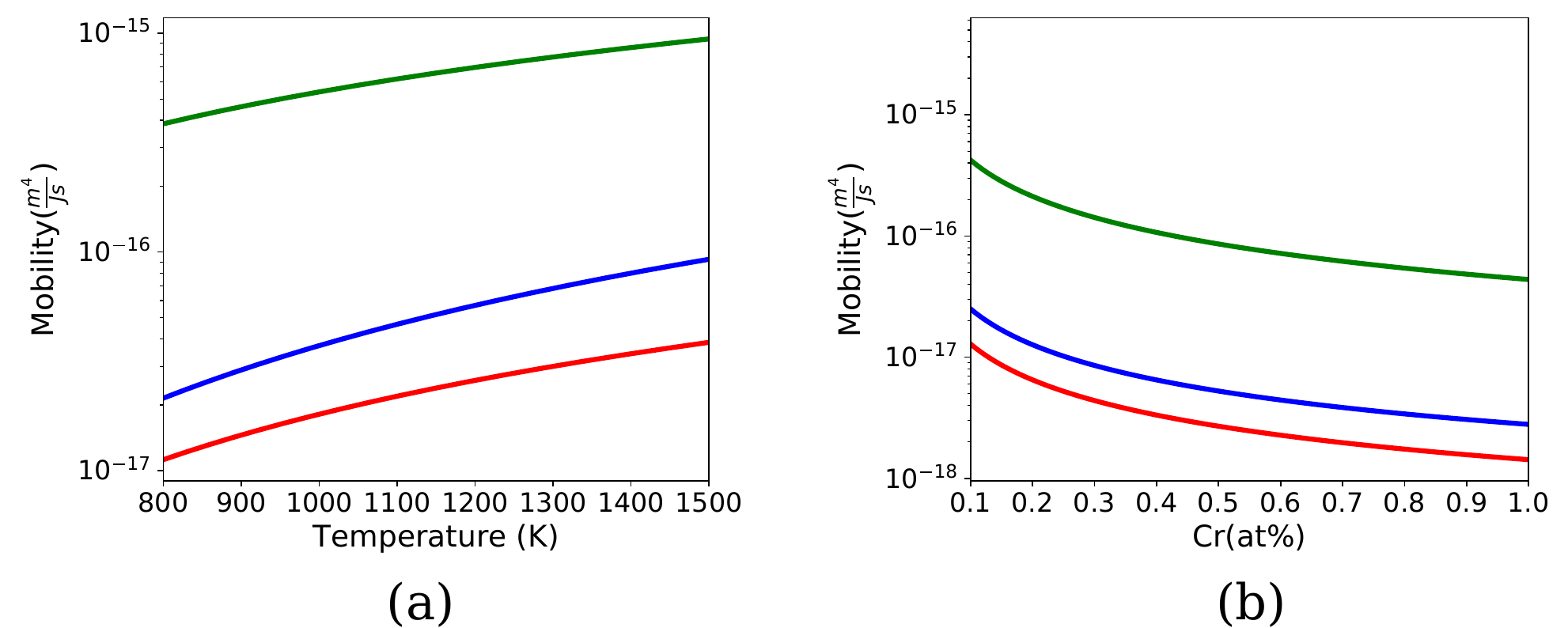}}
\caption{Mobility of the three STGBs, from top to bottom: $\sum 3 (11\bar{2})$, $\sum 9 (\bar{2}21)$, and $\sum 3 (\bar{1}11)$). Variation of mobility as a function of: (a) temperature (at a constant bulk Cr concentration of 0.1\%) and (b) bulk Cr concentration (at a constant temperature of 850 K). Similar plots are given for the rest of the solute atoms in Figures S3-S10, SM.}
\label{fig:mobility_Cr}
\end{figure}
%\textcolor{red}{To estimate the GB mobility, we first need to estimate the GB concentration. We use Equation \ref{eq:gb_concentration_1}, which provides GB concentration for each site. The GB concentration $C_{GB}^{j}$ versus temperature is plotted using the multisite model by considering $C_{b}$ to be $0.1\%$, shown in Figure S2. Results show that in all STGBs considered, segregation decreases with increased temperature as expected. Even for a small bulk concentration, the GB concentration at 100K goes as high as $60\%$. With the increase in temperature, the GB concentration in the case of $\sum 3(11\bar{2})$ drops drastically compared to the other two, indicating a lesser GB thickness. As expected from  $E_{seg}$ data, we observe higher GB concentration for Nb, Mn, and  Ni for all three GBs.}  

Using the effective segregation energy $E_0$, we obtain the mobility from Equation~\ref{eq:mobility}. GB mobility at 850 K for different solutes at different boundaries is compared in the right panel of Figure~\ref{fig:eseg_effective_diff}. Irrespective of the type of solute atom, $\sum 3(11\bar{2})$ has the highest mobility among the three different GBs considered here. This result is consistent with the literature~\cite{TODACARABALLO20121116}, where it was observed that low-energy GB usually has a higher mobility. 

One can explain this trend based on lower solute segregation at $\sum 3(11\bar{2})$ GB [see section~\ref{subsecSE} and Figure S2, SM]. A lower solute segregation implies a weaker solute drag effect, leading to higher mobility of the low-energy $\sum 3(11\bar{2})$ boundary. Overall, solutes having a high magnitude of $E_{seg}$ and low diffusivity (for example Nb) tend to have lesser GB mobility [Figure~\ref{fig:eseg_effective_diff}]. On the other hand, solutes having a low magnitude of $E_{seg}$ and high diffusivity (for example, V) tend to have higher GB mobility. A combination of low $E_{seg}$ and high diffusivity implies less solute segregation at the boundary, leading to a weaker solute drag effect and higher GB mobility. 

\begin{figure}[ht]
\centering 
\framebox{\includegraphics[width=0.7\columnwidth]{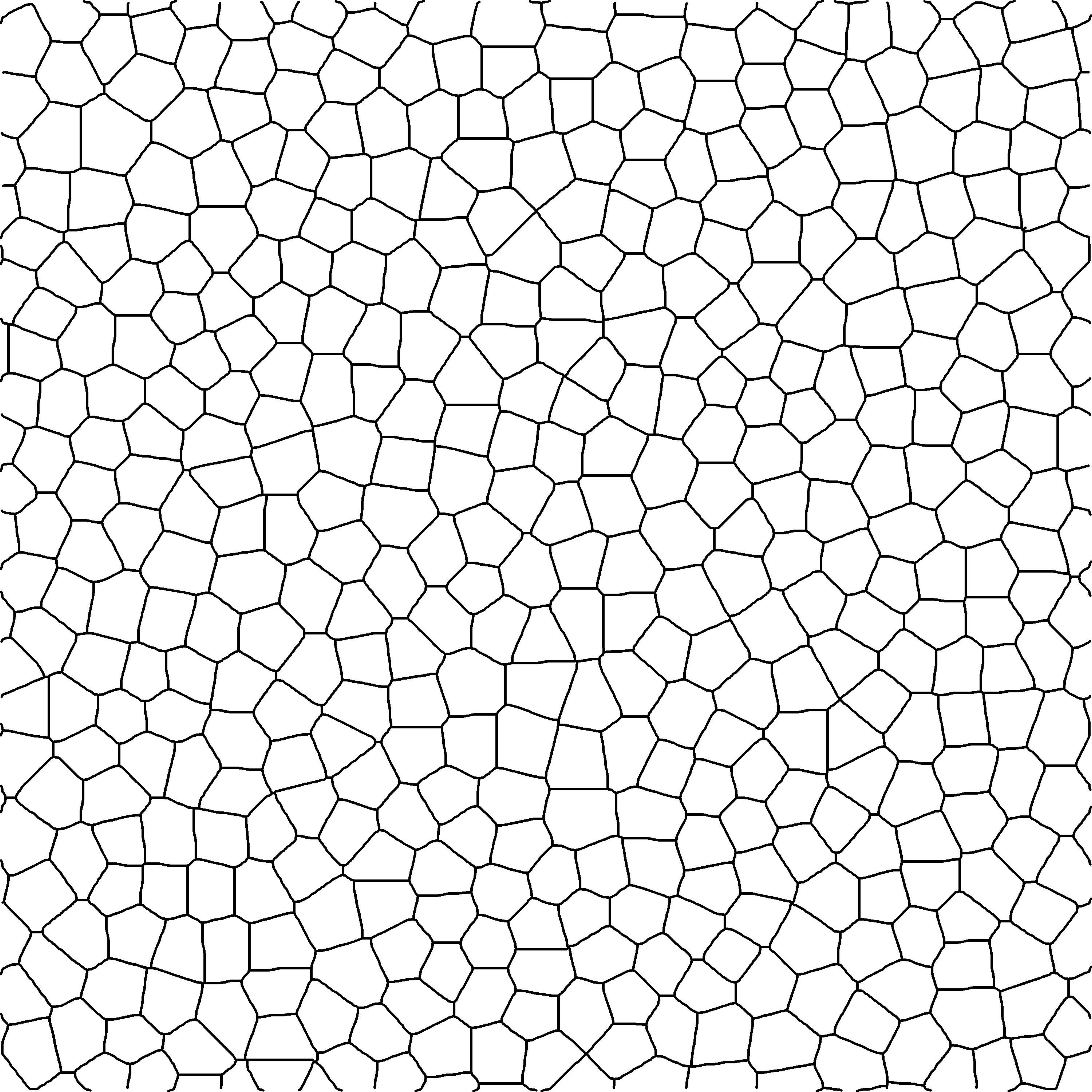}}
\includegraphics[width=0.7\columnwidth]{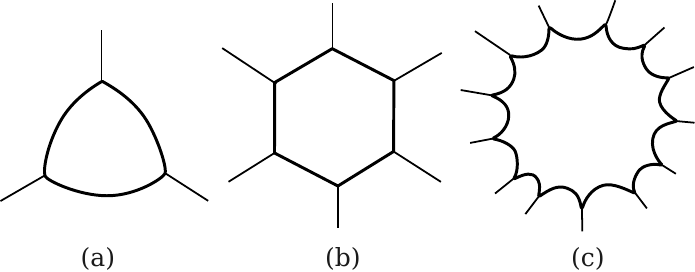}
\caption{(Top) Initial equiaxed microstructure with nearly equal-sized grains generated from Voronoi tessellation. (Bottom) Different grain shapes: (a) Grain with fewer than six sides. Such grains will disappear in later time steps. (b) Six-sided stable grains. (c) Crown-type morphology with a very large number of sides. Such a morphology is observed when a very large grain is surrounded by several small grains, resulting in abnormal grain growth. Note that the initial microstructure mainly has 5 to 7-sided grains, most of which are 6-sided.}
\label{fig:intial_microstructure}
\end{figure}

GB mobility also depends on temperature and bulk concentration of the solute atoms. Considering Cr as the solute atom, the temperature and concentration dependence of GB mobility are illustrated in Figure~\ref{fig:mobility_Cr}. Since solute segregation at the boundary decreases with increasing temperature [Figure S2, SM], the solute drag effect weakens, leading to increased GB mobility at higher temperature [Figure~\ref{fig:mobility_Cr} (a)]. Increasing the overall concentration of the solute (in bulk) also enhances the quantity of solute segregated at the grain boundary, leading to more substantial solute drag effect and decreased GB mobility [Figure~\ref{fig:mobility_Cr} (b)]. A similar trend is observed for the rest of the solute atoms [Figures S3 to S10, SM]. The difference between the mobility of low and high-energy GB persists at all temperatures and bulk compositions of the solute. We take the temperature and overall bulk composition of the solute to be 850 K and 0.1\%, respectively, for subsequent phase-field simulations.

\subsection{Grain growth with mixed grain boundaries}
We now use the mobilities obtained from the \textit{ab initio} calculations and CLS model [Equation~\ref{eq:mobility}] to study the anisotropic grain growth in the presence of solute segregation. We first generate an initial microstructure using Voronoi tesselation, as shown in Figure~\ref{fig:intial_microstructure}. It has mainly five to seven-sided grains, the majority being six-sided. We ensure that all grains have nearly equal diameters, mimicking the grain growth from an equiaxed grain. This model uses a common grain boundary width ($l_{GB}$). As discussed previously, based on our \textit{ab initio} results, we choose $l_{GB}= 1$ nm, which is also consistent with the experimental findings~\cite{Fultz1995}. We consider two cases: one where both types of GBs have low mobility and another where one type has low mobility while the other has high mobility.

%Throughout the study, we scale the mobility of a given microstructure with the highest mobility we are considering for that microstructure.

\begin{figure*}
\centering 
\includegraphics[width=0.87\linewidth]{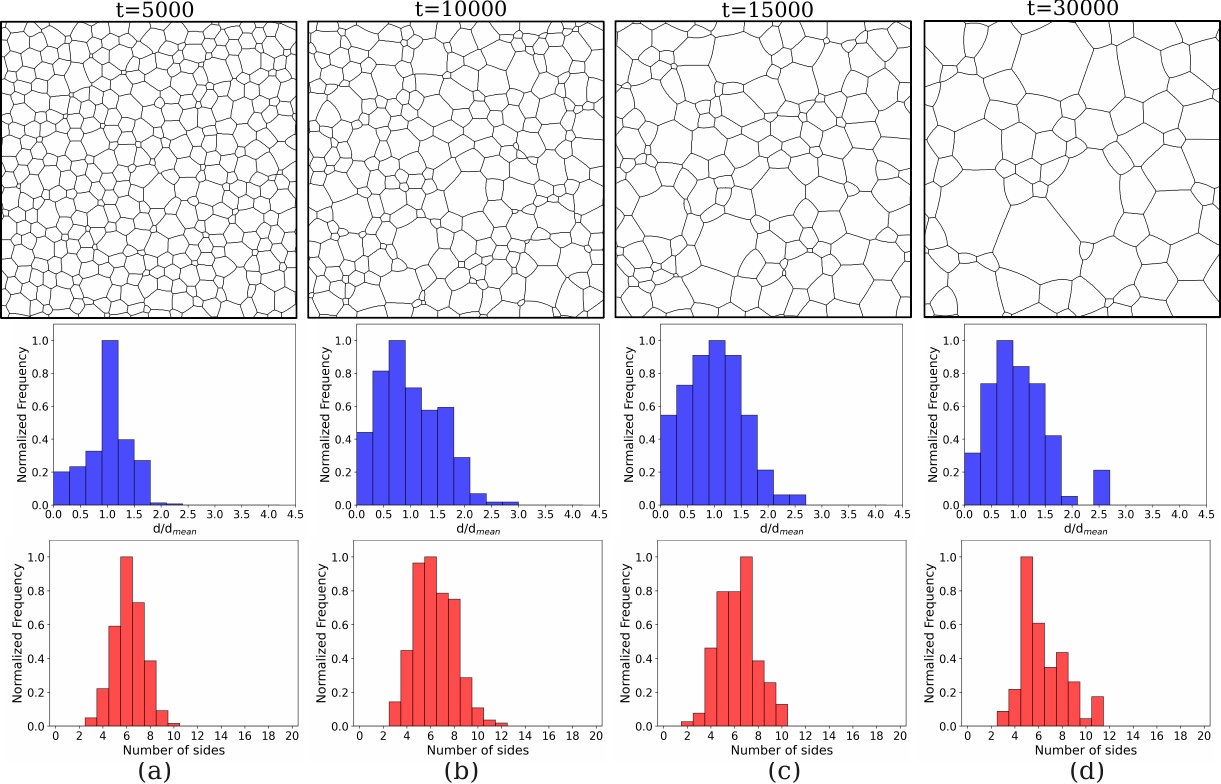}
\caption{Grain growth with mixed boundaries having similar mobility: 10\% $\sum 3 (\bar{1}11)$ and 90\% $\sum 9 (\bar{2}21)$. The top row presents the microstructure evolution from phase-field simulations. The middle row shows the grain size distribution, where $d$ is the grain diameter and $d_{mean}$ is the mean diameter. The bottom row displays the distribution of $n-$sided grains in each microstructure. Note the unimodal grain size distribution, mostly 5 to 8-sided grains.}
\label{fig:low_low_gbs}
\end{figure*}

In the first case, we consider two low-mobility grain boundaries: 10\% of $\sum 3 (\bar{1}11)$  and 90\% of $\sum 9 (\bar{2}21)$ [Figure~\ref{fig:low_low_gbs}]. For the second case, we use a combination of 10\% $\sum 3 (11\bar{2})$ (high mobility) and  90\% $\sum 9 (\bar{2}21)$ (low mobility) grain boundary [Figure~\ref{fig:low_high_gbs}]\textcolor{black}{(see the supplementary video:  link given in Appendix A)}. During the initial time steps [Figure~\ref{fig:low_low_gbs}(a) and Figure~\ref{fig:low_high_gbs}(a)], we do not observe any noticeable change. Both microstructures have grains ranging from 4 to 9 sides. However, as the simulation progresses, differences between the two cases become apparent. If there is a large mobility difference between the GBs, the number of sides in some grains increases significantly. As a result, a bimodal type of distribution starts to appear [Figure~\ref{fig:low_high_gbs} (b)], which is an indication of the onset of abnormal grain growth (AGG). This trend is not observed if both the grain boundaries have low mobility [Figure~\ref{fig:low_low_gbs} (b)]. 

Comparing Figures~\ref{fig:low_low_gbs}(c) and \ref{fig:low_high_gbs}(c), one can clearly identify the effect of AGG in the second case, manifested in the form of a bimodal grain size distribution, whereas the first case still exhibits a unimodal distribution. Furthermore, in the second case, some grains develop a crown-like morphology [Figure~\ref{fig:intial_microstructure} (c)] due to the high-mobility GBs surrounding low-mobility grains, resulting in an inward curvature, which accelerates the GB velocity. In contrast, this crown-like morphology is absent in the first case, leading to normal grain growth. At the final time step, the microstructure exhibiting uniform grain growth is illustrated in Figure~\ref{fig:low_low_gbs}(d), still having a unimodal grain size distribution. On the other hand, the microstructure showing abnormal grain growth [Figure~\ref{fig:low_high_gbs}(d)] has several grains with a large size, having more than nine sides. This study clearly illustrates that a large mobility difference between GBs leads to abnormal grain growth, while GBs with similar mobility lead to more uniform grain growth.

\begin{figure*}
\centering 
\includegraphics[width=0.87\linewidth]{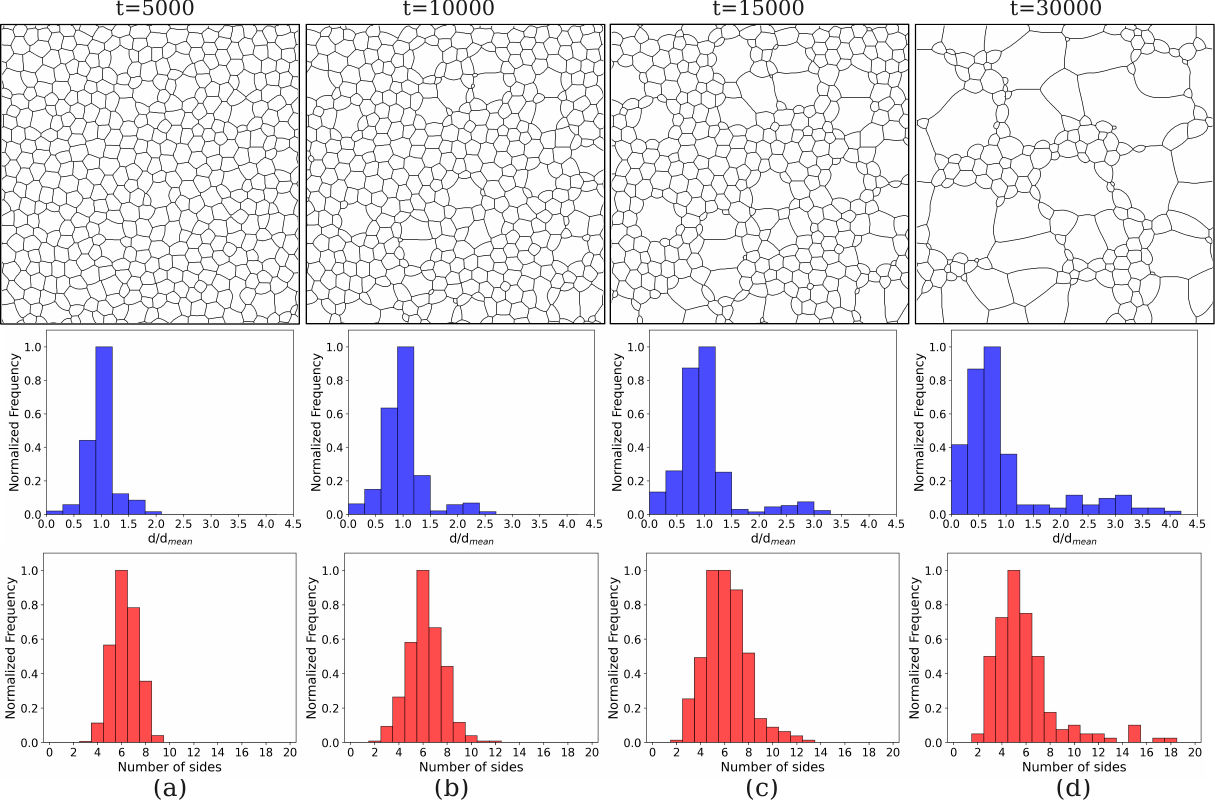}
\caption{Grain growth with mixed boundaries having large mobility difference: 10\% high mobility $\sum 3 (11\bar{2})$ and 90\% low mobility $\sum 9 (\bar{2}21)$(90\%). The figures are arranged similarly to Figure~\ref{fig:low_low_gbs}. Note the bimodal grain size distribution, with a significant number of more than 9-sided grains.} 
\label{fig:low_high_gbs}
\end{figure*}

\textcolor{black}{To understand the influence of GB energy and GB mobility independently on microstructural evolution, we consider three hypothetical cases. We take the high mobility, low energy GB $\Sigma 3\ [110](11\bar{2})$ and low mobility, high energy GB $\Sigma 9\ [110](\bar{2}21)$ in a $10:90$ ratio and conduct the following phase-field simulations. 
\begin{itemize}
\item 
Case 1: Both GBs are assigned the same energy (to that of $\Sigma 9\ [110](\bar{2}21)$), but their actual mobilities are used. 
\item
Case 2: Both GBs are given the same mobility (to that of $\Sigma 9\ [110](\bar{2}21)$), but their actual energies are used. 
\item3
Case 3: Both GBs retain their actual values for both mobility and energy.
\end{itemize}
Interestingly, case 1 resulted in normal grain growth [Figure S11(a), SM]. On the other hand, cases 2 and 3 resulted in AGG [Figure S11(b)-(c), SM]. This comparison clearly indicates that GB mobility plays a more dominant role than GB energy in  triggering the AGG behavior observed in the system.}

\begin{figure}[ht]
\centering 
\fcolorbox{black}{white}{\includegraphics[width=1.0\linewidth]{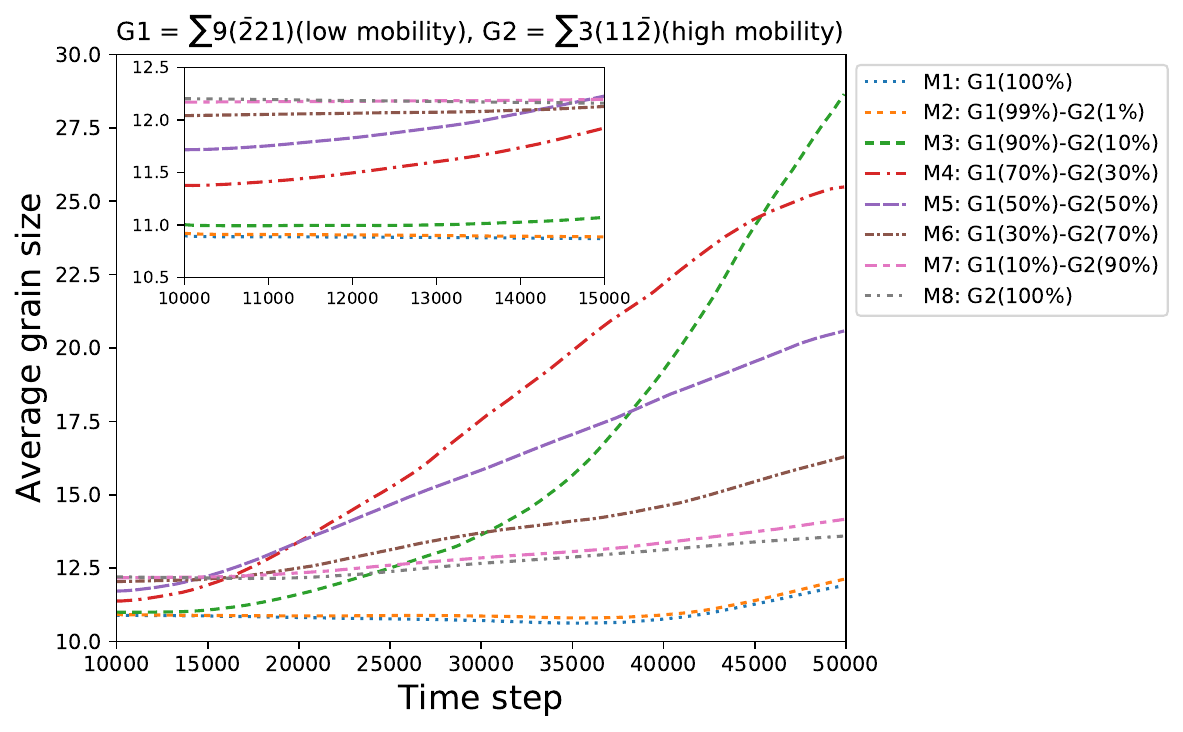}}
\caption{Comparison of grain growth with different fractions of high and low mobility GB. Abnormal grain growth does not occur when we start with 100\% high mobility ($M_8$) or 100\% low mobility ($M_1$) GB. Abnormal grain growth is more prominent when the fraction of high mobility GB is relatively low, such as $M_3$, followed by $M_4$ and $M_5$. In case of $M_2$, the fraction of high mobility boundary is too low, while in case of $M_6, M_7$, the fraction of high mobility boundary is too high for abnormal grain growth to take place.}
\label{fig:avg_grain_size_mixed_gbs}
\end{figure}

While we have established that a combination of 10\% $\sum 3 (11\bar{2})$ (high mobility) and  90\% $\sum 9 (\bar{2}21)$ (low mobility) grain boundary leads to abnormal grain growth, it is unlikely that mixing them in any proportion will lead to similar outcome. We further investigate the grain growth behavior after mixing the low and high mobility boundaries in different proportions, from M1 to M8 [see Figure~\ref{fig:avg_grain_size_mixed_gbs}], \textcolor{black}{while the microstructures corresponding to mixture M4, M5, and M6 are presented in Figures S12–S14 of the supplementary material.} At 10000 time steps, we observe the following average grain size trend: $M_8 > M_7 > M_6 > M_5 > M_4 > M_3 > M_2 > M_1$. This trend is natural because $M_1$ is 100\% low mobility GB, while $M_8$ is 100\% high mobility GB, and the rest are a mixture of them, with the proportion of high mobility GB gradually increasing from $M_2$ to $M_7$.

As we progress in time, the trend mentioned above changes. At about 15000 time step, $M_5$ (50\% high mobility) and $M_7$ (90\% high mobility) has higher average grain size compared to $M_8$ (100\% high mobility) and they can be arranged as $M_5 > M_7 > M_8 > M_6 > M_4 > M_3 > M_2 > M_1$. With further increase in simulation time, we observe that after 25000 time steps, $M_4$ (30\% high mobility) has the highest average grain size, and the order changes to $M_4 > M_5 > M_6 > M_7 > M_8 > M_3 > M_2 > M_1$. Finally, after 50000 time steps, $M_3$ (10\% high mobility) has the highest grain average grain size, and the order changes to $M_3 > M_4 > M_5 > M_6 > M_7 > M_8 > M_2 > M_1$. 

Based on Figure~\ref{fig:avg_grain_size_mixed_gbs} and related text (previous paragraph), it is clear that abnormal grain growth occurs when we start with a small amount of high mobility GB. In our study, $M_3$ (combination of 10\% high and 90\% low mobility GB) shows the highest rate of average grain size increase at the later time stage [Figure~\ref{fig:avg_grain_size_mixed_gbs}], leading to abnormal grain growth. As the fraction of high mobility grain boundaries increases, the tendency of abnormal grain growth decreases. One can explain this trend based on the following observation. When the fraction of high mobility GB is higher, the grains with high mobility GB are not surrounded by a sufficient number of low mobility GB. As a result, there is less chance of crown-type morphology formation, a crucial factor for abnormal grain growth. However, the fraction of high mobility GB must have more than some critical value for abnormal grain growth to take place. For example, $M_7$ (combination of 1\% high and 99\% low mobility) does not show any abnormal grain growth, and it is very similar to $M_8$ (100\% low mobility) [Figure~\ref{fig:avg_grain_size_mixed_gbs}]. When the fraction of high mobility GB is too small, such GBs are sparsely scattered and can not form a closed loop.
 
\begin{figure}[ht]
\centering 
\includegraphics[width=1.0\linewidth]{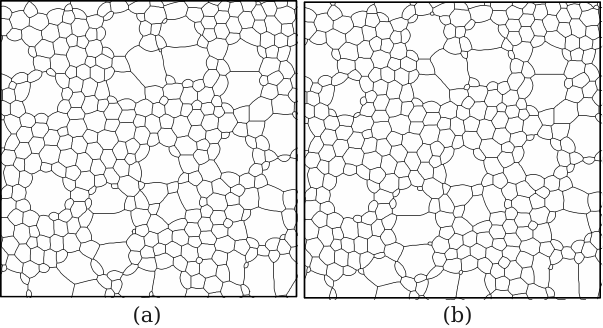}
\caption{Microstructures at the $15,000^{th}$ time step for a mixture of: (a) $\sum 3 (11\bar{2})$ (10\%), and $\sum 9 (\bar{2}21)$ (90\%) boundaries; (b) $\sum 3 (11\bar{2})$ (10\%), $\sum 9 (\bar{2}21)$ (45\%), and $\sum 3 (\bar{1}11)$ (45\%) boundaries.} 
\label{fig:comparison_high_low}
\end{figure}

It is important to understand that the main factor behind abnormal grain growth is the anisotropy of grain boundary mobility, with the high mobility boundary present in a lesser proportion than the low mobility boundary. To verify, we run two simulations. In both, we keep the proportion of the high mobility $\sum 3 (11\bar{2})$ boundary fixed to 10\%. In one simulation, we take the low mobility boundary to be $\sum 9 (\bar{2}21)$ (90\%). The other simulation runs with a mixture of low mobility boundaries; $\sum 3 (\bar{1}11)$ (45\%) and $\sum 9 (\bar{2}21)$  (45\%). As shown in Figure~\ref{fig:comparison_high_low}, both microstructures are nearly identical, as the abnormal grain growth behavior in both cases follows a similar trend. This is unsurprising as both the low mobility boundaries of $\alpha-$Fe with Cr solute have comparable GB mobility, significantly smaller than the high mobility boundary [Figure~\ref{fig:eseg_effective_diff}]. Thus, the type of grain boundaries present is not so relevant for abnormal grain growth. What matters is the anisotropy in terms of GB mobility, and the high mobility grain boundary should be present in a lesser proportion than the low mobility boundary.  

\begin{figure*}[htpb]
\centering 
\includegraphics[width=1.0\linewidth]{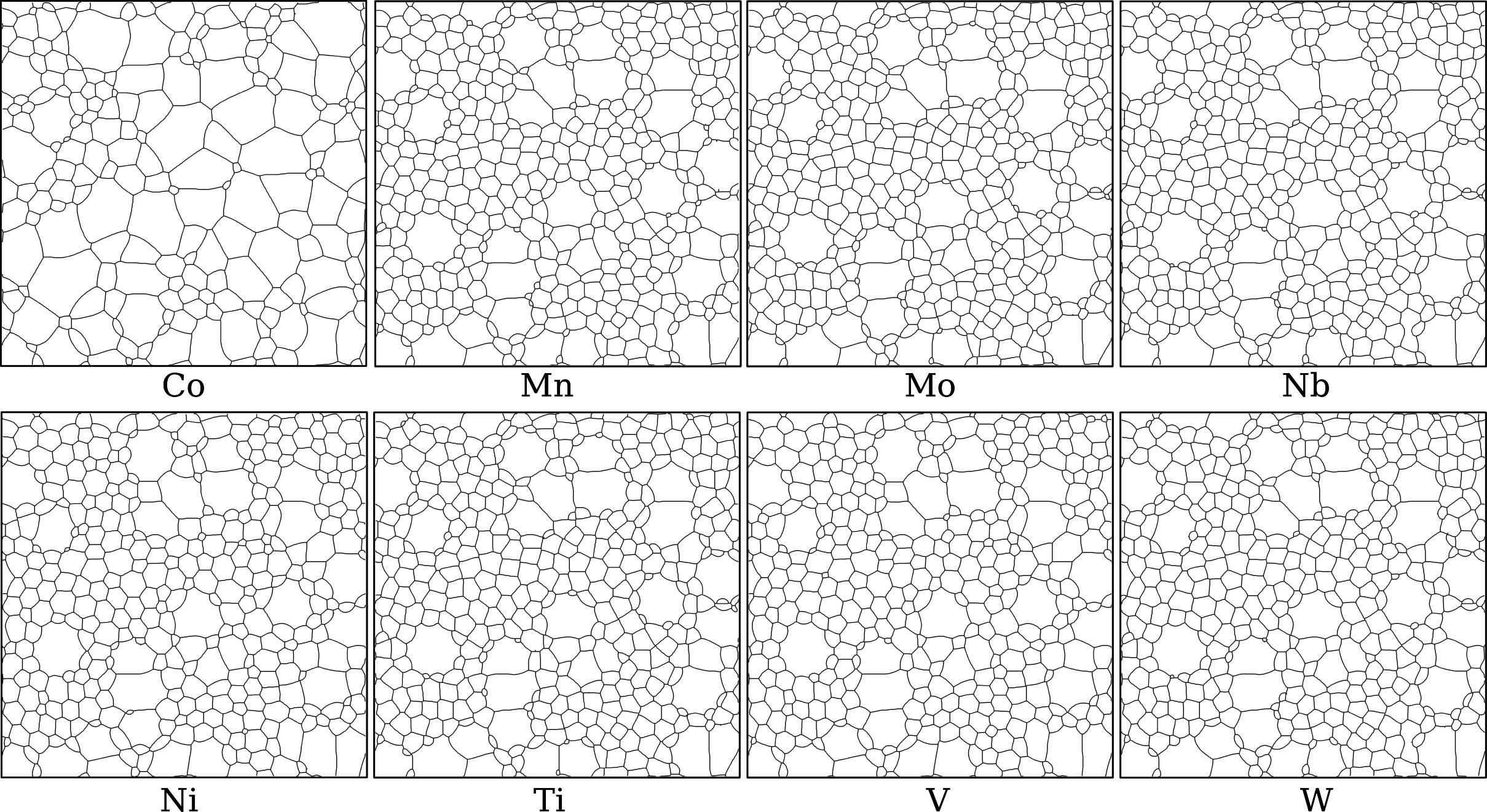}
\caption{Comparison of simulated microstructures of $\alpha-$Fe containing different solute atoms. In all the cases, the initial microstructure contains a mixture of three types of GBs: 10\% $\sum 3 (11\bar{2})$, 45\% $\sum 9 (\bar{2}21)$, and 45\% $\sum 3 (\bar{1}11)$.} 
\label{fig:low_high_gbs_all}
\end{figure*}

Finally, we do grain growth studies for each of the solute atoms, starting with a mixture of 10\% $\sum 3 (11\bar{2})$, 45\% $\sum 9 (\bar{2}21)$, and 45\% $\sum 3 (\bar{1}11)$ GB. For direct comparison, mobilities are scaled by the value of $\sum 3 (11\bar{2})$ for each of the solutes. As shown in Figure~\ref{fig:low_high_gbs_all}, the morphology of the microstructures for all the solutes is qualitatively similar, except for Co. The similarity proves that, whatever the segregating element, abnormal grain growth would still be observed when the high mobility GBs are present in small proportions. In the case of Co, $\sum 9 (\bar{2}21)$ GB (which has low mobility in the case of other solutes) has high mobility, comparable to the $\sum 3 (11\bar{2})$ [Figure~\ref{fig:eseg_effective_diff}]. As a result, the total amount of high mobility boundaries in the case of Co segregation is 55\%, which is too high for the abnormal grain growth to start. 

\subsection{Experimental evidence of AGG behaviour}
There have been several instances where AGG has been observed in $\alpha$-Fe and its alloys. For example, annealed steel shows the highest presence of $\sum 3$ boundaries in the microstructure compared to all other CSL boundaries~\cite{Beladi2012}. In Invar Fe-36\%Ni, it has been experimentally observed that the area of $\sum 3$ boundaries increased significantly with the increase in temperature and a high distribution of twin boundaries has been reported at around 1123 K~\cite{He2018}.

\textcolor{black}{It has been established that plastically deformed pure $\alpha$-Fe can develop a $\langle 110 \rangle$ fiber texture. Following annealing for approximately eight hours, this texture remains dominant, accounting for nearly 40\% of the microstructure~\cite{JANICKI2001187}. Annealing at 873~K for a few hours has been shown to result in anisotropic grain growth, indicating that such anisotropy may arise from the presence of $\langle 110 \rangle$-oriented grains. However, AGG was not observed in them, particularly due to the absence of any solute drag effect and low temperature.}

\textcolor{black}{Recently, AGG has been reported in thin specimens of commercially pure iron containing trace amounts ($\approx$0.1\%) of Mn, Cr, and Ni~\cite{ALMEIDAJR202011099}. In this study, the material was wire-drawn along the $\langle 110 \rangle$ direction, inducing a strong $\langle 110 \rangle$ fiber texture. Upon annealing, the microstructure preserved this preferred orientation, and AGG was observed specifically at the wire center, where equiaxed grains dominated. Although AGG was reported at an annealing temperature of 1123~K, the authors claimed that its occurrence can also happen at even lower temperatures. Overall, they attributed the primary driving force for AGG to be the influence of crystallographic texture, which is also shown in our simulation results. This indicates that the combined effect of the pronounced difference in GB mobility and solute drag effect can be one of the possible reasons for promoting the AGG behavior.}

\section{Conclusions}

We have performed a multiscale study of grain growth in STGB of $\alpha$-Fe for three $\langle 110 \rangle$ axis type GBs ($\sum 3 (11\bar{2})$,  $\sum 9 (\bar{2}21)$ and $\sum 3 (\bar{1}11)$) by using a multiscale framework combining the \textit{ab initio} calculations and phase-field simulations. The following key conclusions can be drawn from this study.

\begin{enumerate}
\item We obtain segregation energies from \textit{ab initio} calculations for nine types of transition elements in bulk $\alpha$-Fe. Based on the segregation values calculated, we get the trend for segregation tendency, with Nb (V) being the most (least) prominent. 
\item Solute excess is obtained from the GB concentration, which is utilized to obtain GB mobilites. For all types of solute segregation, the low energy $\sum 3 (11\bar{2})$ possesses the highest mobility, indicating the lowest solute drag effect.
\item The presence of only high or only low mobility boundaries leads to normal grain growth. For the onset of AGG, a mixture of high ($\sum 3 (11\bar{2})$) and low mobility ($\sum 9 (\bar{2}21)$/$\sum 3 (\bar{1}11)$) GB is necessary, but not a sufficient condition.
\item Distribution of GB types is also very significant for AGG. Suppose the fraction of high mobility GBs is too low. In that case, such grains are so sparsely distributed that the microstructure fails to form the crown-like morphology, ultimately leading to normal grain growth. When 10-30\% of the GBs are high-mobility type, crown-like morphologies are formed, leading to AGG.   
\end{enumerate}

\section{Acknowledgements} RM and SB acknowledges financial support from SERB core research grant (CRG/2019/006961). We express our gratitude to the National Super Computing Mission (NSM) for granting access to the computing resources of ``PARAM Sanganak'' at IIT Kanpur, which is managed by CDAC and funded by the Ministry of Electronics and Information Technology (MeitY) and the Department of Science and Technology (DST), Government of India. We also thank the ICME National Hub and the Central Computing facility at IIT Kanpur for their support in providing high-performance computing resources.

\appendix

%\section{}
%A supporting video will be included in this appendix as supplementary information and added at a later stage (now it is uploaded separately).

%\bibliography{bibfile}

\end{document}